\documentclass[conference,9pt]{IEEEtran}
\IEEEoverridecommandlockouts
\usepackage{cite}
\usepackage{amsmath,amssymb,amsfonts}
\usepackage{makecell}
\usepackage{graphicx}
\usepackage{textcomp}
\usepackage{xcolor}
\usepackage{enumitem}
\usepackage{multirow,multicol}
\usepackage{algorithm}
\usepackage[noend]{algpseudocode}
\usepackage[a4paper, total={184mm,239mm}]{geometry}
\usepackage{hyperref}
\usepackage{booktabs}
\def\BibTeX{{\rm B\kern-.05em{\sc i\kern-.025em b}\kern-.08em
    T\kern-.1667em\lower.7ex\hbox{E}\kern-.125emX}}

\usepackage{array}
\newcolumntype{P}[1]{>{\centering\arraybackslash}p{#1}}

\usepackage{tikz}
\newcommand{\blackcircle}[1]{%
  \tikz[baseline=(char.base)]{
    \node[shape=circle,draw,fill=black,text=white,inner sep=0.2pt,font=\footnotesize] (char) {#1};
  }%
}

\newcommand{\squishlist}{
 \begin{list}{$\bullet$}
  { \setlength{\itemsep}{0pt}
     \setlength{\parsep}{3pt}
     \setlength{\topsep}{3pt}
     \setlength{\partopsep}{0pt}
     \setlength{\leftmargin}{1.5em}
     \setlength{\labelwidth}{1em}
     \setlength{\labelsep}{0.5em} } }
\newcommand{\squishend}{
  \end{list}  }

\newcommand{\papername}[1]{AC-Refiner}
    
\begin{document}


\title{AC-Refiner: Efficient Arithmetic Circuit Optimization Using Conditional Diffusion Models}

\author{
    \IEEEauthorblockN{
        Chenhao Xue\textsuperscript{1},
        Kezhi Li\textsuperscript{2},
        Jiaxing Zhang\textsuperscript{1}, 
        Yi Ren\textsuperscript{1,3}, 
        Zhengyuan Shi\textsuperscript{1}, \\
        Chen Zhang\textsuperscript{4},
        Yibo Lin\textsuperscript{1,5,6},
        Lining Zhang\textsuperscript{7},
        Qiang Xu\textsuperscript{2,8},
        Guangyu Sun\textsuperscript{1,5,6,*}
        \thanks{\textsuperscript{*}Corresponding author: Guangyu Sun (\href{mailto:gsun@pku.edu.cn}{gsun@pku.edu.cn})}
    }
    \IEEEauthorblockA{
        \textsuperscript{1}\textit{School of Integrated Circuits, Peking University}, Beijing, China \\
        \textsuperscript{2}\textit{Department of Computer Science and Engineering, The Chinese University of Hong Kong}, Sha Tin, Hong Kong S.A.R. \\
        \textsuperscript{3}\textit{School of Software and Microelectronics, Peking University}, Beijing, China \\
        \textsuperscript{4}\textit{Shanghai Jiao Tong University}, Shanghai, China \\
        \textsuperscript{5}\textit{Institute of Electronic Design Automation, Peking University}, Wuxi, China \\
        \textsuperscript{6}\textit{Beijing Advanced Innovation Center for Integrated Circuits}, Beijing, China \\
        \textsuperscript{7}\textit{School of Electronic and Computer Engineering, Peking University}, Shenzhen, China \\
        \textsuperscript{8}\textit{National Center of Technology Innovation for EDA}, Nanjing, China \\
        \{xch927027, yibolin, eelnzhang, gsun\}@pku.edu.cn, \{yiren20,zjx\}@stu.pku.edu.cn \\
        \{kzli24, zyshi21, qxu\}@cse.cuhk.edu.hk, chenzhang.sjtu@sjtu.edu.cn
    }
}

\maketitle

\begin{abstract}

Arithmetic circuits, such as adders and multipliers, are fundamental components of digital systems, directly impacting the performance, power efficiency, and area footprint.
However, optimizing these circuits remains challenging due to the vast design space and complex physical constraints. While recent deep learning-based approaches have shown promise, they struggle to \textit{consistently explore high-potential design variants}, limiting their optimization efficiency. 
To address this challenge, we propose \textit{\papername{}}, a novel arithmetic circuit optimization framework leveraging conditional diffusion models. 
Our key insight is to reframe arithmetic circuit synthesis as a conditional image generation task. By carefully conditioning the denoising diffusion process on target quality-of-results (QoRs), \papername{} consistently produces high-quality circuit designs. Furthermore, the explored designs are used to fine-tune the diffusion model, which focuses the exploration near the Pareto frontier.
Experimental results demonstrate that \papername{} generates designs with superior Pareto optimality, outperforming state-of-the-art baselines. The performance gain is further validated by integrating \papername{} into practical applications.

\end{abstract}


\section{Introduction}

In digital circuit designs, arithmetic circuits—such as adders and multipliers—serve as fundamental building blocks. These components play a critical role in determining the performance, power consumption, and area efficiency of hardware systems, particularly in computation-intensive applications like digital signal processing and neural networks. As a result, optimizing arithmetic circuits is essential for achieving high-performance and energy-efficient systems.

The optimization of arithmetic circuits, however, presents significant challenges due to the vast design space, which grows exponentially with input bit width.
Over the years, a range of manual design techniques~\cite{sklansky1960conditional,brent1982regular,kogge1973parallel,wallace1964suggestion,dadda1990some} and algorithmic methods~\cite{liu2007optimum,matsunaga2007area,roy2013towards,moto2018prefix,lin2024size,kumm2014pipelined,xiao2021gomil,zuo2024ufo}, which focus on minimizing theoretical properties such as logic depth and size.
Despite these efforts, a substantial gap remains between theoretical optimization and real-world performance. This discrepancy stems from the inherent difficulty of accurately modeling physical implementation constraints (e.g. capacitive loading and routing congestion) at early design stages~\cite{ma2018cross,roy2021prefixrl}. Consequently, purely theoretical optimizations cannot guarantee optimal quality-of-results (QoRs) in practice.

Recent advancements have introduced learning-based approaches to address the limitations of theoretical optimization, facilitating more effective design space exploration (DSE) of adders~\cite{roy2021prefixrl,lai2024scalable,geng2021high,song2024circuitvae} and multipliers~\cite{zuo2023rl,lai2024scalable,feng2024gomarl} for physically optimized implementations. Nevertheless, existing learning-based methods still face fundamental limitations in optimization efficiency, particularly when scaling to large and complex arithmetic units such as high-performance multipliers.

On the one hand, reinforcement learning (RL) has been applied to optimize designs through iterative interactions with physical synthesis tools~\cite{roy2021prefixrl,zuo2023rl,feng2024gomarl,lai2024scalable}. While RL-based methods demonstrate potential, they often require extensive trial-and-error before converging on optimized solutions. This iterative process can be prohibitively time-consuming and computationally expensive, particularly for large-scale multipliers~\cite{lai2024scalable}. Consequently, RL-agents often struggle to effectively navigate the vast design space within limited budgets, yielding suboptimal design quality.

On the other hand, generative models offer a promising alternative to address the efficiency limitation of RL. By capturing the structural patterns of prior arithmetic circuit designs, they can rapidly sample diverse promising candidates for parallel evaluation. Prior research has explored variational autoencoders (VAEs) for prefix adder optimization~\cite{geng2021high,song2024circuitvae}. However, VAEs exhibit limited capability to model fine-grained structural details~\cite{gulrajani2016pixelvae}.
For arithmetic modules with complex design constraints (e.g. multiplier compressor trees), it is difficult for VAEs to sample valid designs.

\begin{figure}
    \centering
    \includegraphics[width=\linewidth]{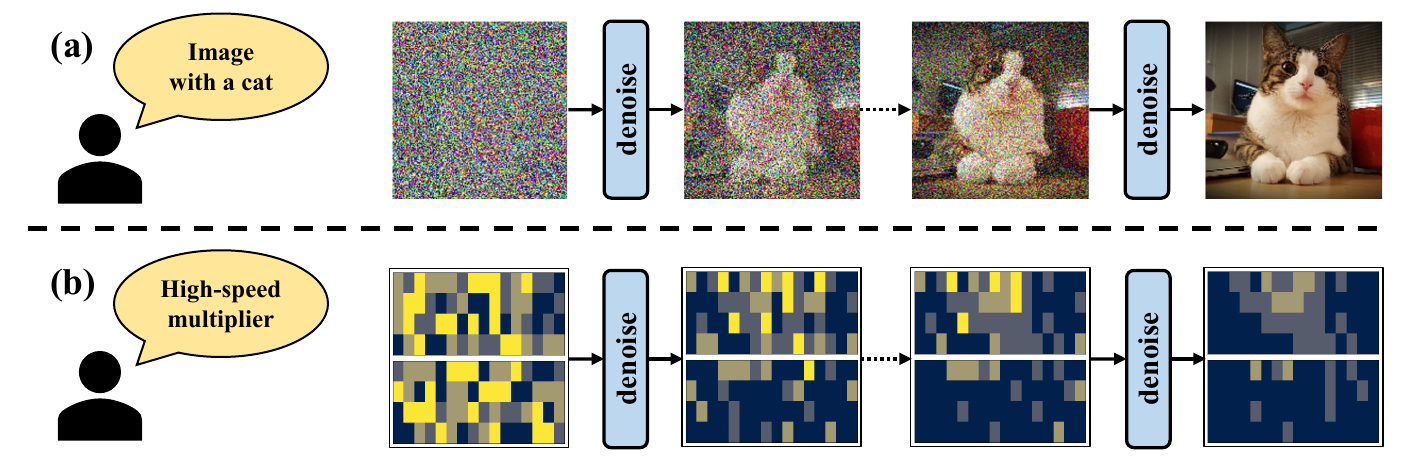}
    \caption{(a) Image generation with a conditional diffusion model. (b) \papername{} represents arithmetic circuits similar to images, and conditionally samples circuit designs with desired properties, enabling efficient optimization.}
    \label{fig:analogy}
\end{figure}

In this paper, we aim to address the limitations of existing learning-based arithmetic circuit optimization by leveraging recent advancements of generative models. To this end, we propose \textbf{\papername{}}, which employs conditional diffusion models~\cite{sohl2015deep,ho2020denoising,song2020denoising} for the comprehensive and efficient optimization of multipliers, as shown in Fig.~\ref{fig:analogy}. The key contributions of \papername{} are summarized as follows:
\begin{itemize}
    \item \textbf{Circuit Generation:} Unlike VAEs, which generate arithmetic circuit designs in a single inference step, diffusion models employ an \textit{iterative inference process}. Starting from a random initial design, they progressively refine the structure, ensuring close adherence to design constraints. Incorporated with proper legalization procedures, diffusion models allow \papername{} to deliver valid and diverse candidate designs.
    \item \textbf{Guided Optimization:} To achieve efficient optimization, \papername{} introduces \textit{conditional sampling mechanism} to explore high-potential designs. Specifically, \papername{} adopts neural cost predictors whose gradient steer the inference process toward promising outcomes. Additionally, \papername{} leverages discovered high-quality designs to fine-tune the models, which facilitates focused exploration near the Pareto frontier.
\end{itemize}


We apply \papername{} to the comprehensive optimization of multipliers, covering both compressor trees and carry-propagate adders.
Our experimental results show that \papername{} discovers multipliers that Pareto-dominate all baseline methods at various bit-widths, achieving reductions of up to 15\% in delay and 10\% in area, respectively.
The effectiveness of \papername{} is further validated through implementing multipliers within systolic arrays, highlighting its practical applicability in real-world designs.
We extensively ablate the proposed techniques, including gradient-guided sampling and model fine-tuning, and demonstrate their effectiveness for improving optimization efficiency.

The remainder of this paper is organized as follows:
Section~\ref{sec:background} provides background on multiplier design and diffusion models.
Section~\ref{sec:methodology} details the \papername{} framework.
Section~\ref{sec:experiment} presents the evaluation results.
Finally, Section~\ref{sec:conclusion} concludes the paper.

\section{Background}
\label{sec:background}

\begin{figure}[!t]
  \centering
  \includegraphics[width=0.85\linewidth]{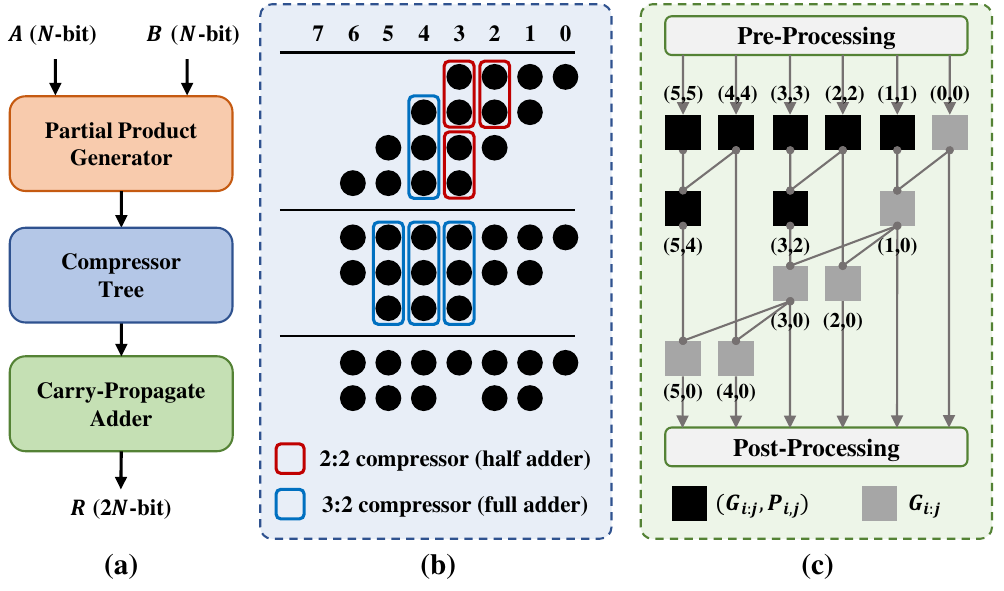}
  \caption{(a) Multiplier architecture. \papername{} targets critical submodules of multipliers, including (b) compressor tree and (c) prefix adder.}
  \label{fig:mult_arch}
\end{figure}

\subsection{Multiplier Architecture}

As shown in Fig.~\ref{fig:mult_arch}(a), a multiplier consists of three main components: a partial product generator (PPG), a compressor tree (CT), and a carry-propagate adder (CPA). 

\squishlist
    \item \textit{Partial Product Generator: } 
    The PPG is responsible for generating partial products (PPs), which are shifted into different columns to represent distinct power-of-two weights.
    A commonly employed \texttt{AND} gate-based PPG generates $N^2$ PPs for an $N$-bit multiplier.

    \item \textit{Compressor Tree: }
    As illustrated in Fig.~\ref{fig:mult_arch}(b), the CT compresses PPs for multiple stages until no more than two PPs remain in each column.
    It is typically composed of various types and quantities of compressors, with full adders (FAs) and half adders (HAs) being predominantly used.
    These basic compressors take multiple bits (3 for FA and 2 for HA) of the same weight as input, producing one bit of the original weight and another of higher weight. 

    \item \textit{Carry-Propagate Adder: } 
    The CPA aggregates the compressed PPs from the CT to produce the final multiplication result.
    To optimize for performance and accommodate the non-uniform signal arrival time~\cite{zuo2024ufo}, \papername{} adopts prefix adders to implement CPA. Prefix adders exhibit a tree-based structure due to the inherent associativity of prefix computation, as shown in Fig.~\ref{fig:mult_arch}(c).
\squishend

\subsection{Diffusion Model}
\label{sec:background_diffusion}

Diffusion models are strong generative models, demonstrating superior performance over VAEs when introduced to controllable generation of high-quality images~\cite{ho2020denoising}.
Due to their exceptional capabilities, diffusion models have been extended to various domains~\cite{li2022diffusion,huang2022fastdiff,wu2024protein,chi2023diffusion,sun2023difusco}.
Conceptually, diffusion models consist of a $T$-step forward process and a $T$-step reverse process. 
In the forward process, the diffusion model gradually destroys clean training data $x_0$ with Gaussian noise for multiple timesteps:
\begin{equation}
    x_t = \sqrt{\alpha_t} \cdot x_{0} + \sqrt{1 - \alpha_t} \cdot \epsilon, \quad \epsilon \sim \mathcal{N}(0, \mathbf{I})
    \label{eq:diffusion_forward}
\end{equation}
where $x_t$ represents the noisy data at timestep $t \in [1, T]$, and $\{\alpha_t\}_{t=1}^{T}$ denotes the noise schedule. In the reverse process, the diffusion model learns to recover $x_0$ by training a neural network $\epsilon_\theta$ to predict the injected noise $\epsilon$:
\begin{equation}
    \epsilon_\theta(x_t, t) \approx \epsilon = \dfrac{x_t - \sqrt{\alpha_t} \cdot x_{0}}{\sqrt{1 - \alpha_{t}}}
    \label{eq:diffusion_reverse}
\end{equation}
At inference time, the diffusion model starts from a randomly sampled noise $x_T \sim \mathcal{N}(0, \mathbf{I})$ and gradually denoises the data using $\epsilon_\theta$. Since the vanilla sampling typically requires $T=1000$ steps~\cite{ho2020denoising}, many accelerated methods are proposed~\cite{song2020denoising,liu2022pseudo}. These methods share a similar formulation, which first predicts the injected noise $\hat{\epsilon} = \epsilon_\theta(x_t, t)$, and produces a less noisy data $x_{t-1}$ w.r.t. $x_t$, $\hat{\epsilon}$, and $t$:
\begin{equation}
   x_{t-1} = \mathcal{S}(x_t, \hat{\epsilon}, t)
    \label{eq:diffusion_sample}
\end{equation}

\section{Methodology}
\label{sec:methodology}

\begin{figure}[!t]
  \centering
  \includegraphics[width=\linewidth]{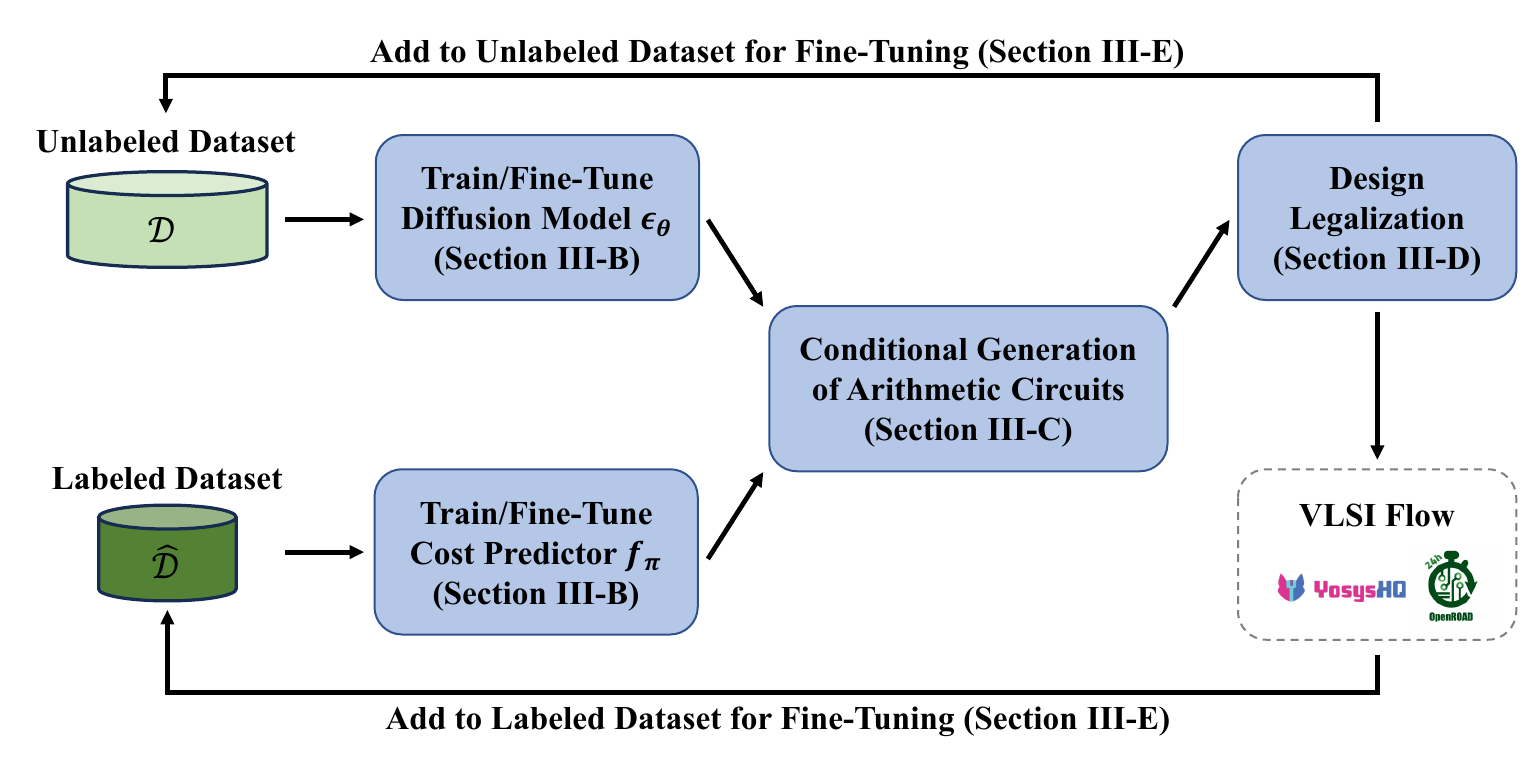}
  \caption{\papername{} framework overview. After training the models, we use gradient-guided sampling to explore high-potential arithmetic circuit designs, which undergo legalization and VLSI synthesis. The explored high-quality designs are leveraged to fine-tune the model for the next optimization round.}
  \label{fig:framework_overview}
\end{figure}

\subsection{Framework Overview}
\label{sec:framework}

In this section, we describe the proposed \papername{} framework for optimizing arithmetic circuits.
As illustrated in Fig.~\ref{fig:framework_overview}, we commence by training diffusion models and cost predictors, where the compressor trees and prefix trees are encoded in image-like binary bitmaps (Section~\ref{sec:training}).
Next, we employ guided sampling to generate promising design candidates (Section~\ref{sec:sampling}), and adopt the legalization procedure to ensure design correctness (Section~\ref{sec:legalization}).
Finally, we leverage the explored high-quality designs to fine-tune the models before proceeding to the next optimization round (Section~\ref{sec:finetuning}).

\subsection{Model Training}
\label{sec:training}

In this section, we first explain how to represent arithmetic circuits with image-like multi-dimensional tensors. Then, we introduce how to train the diffusion models and neural cost predictors.

\begin{figure}[!t]
  \centering
  \includegraphics[width=0.9\linewidth]{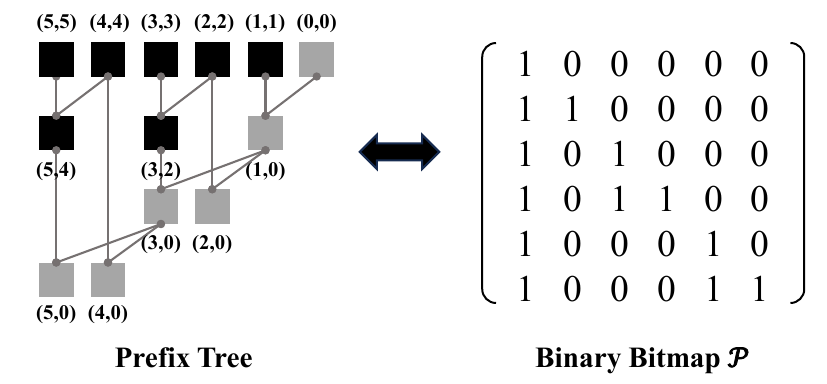}
  \caption{Data representation of prefix adder.}
  \label{fig:prefix_representation}
\end{figure}

\noindent \textbf{Prefix Adder Representation.}
As shown in Fig.\ref{fig:prefix_representation}, we represent $N$-bit prefix adders as bitmap $\mathcal{P} \in \{0, 1\}^{N \times N}$, where $\mathcal{P}[i, j] = 1$ indicates the presence of prefix signals $G_{i,j}$ ($i \geq j$) and $P_{i,j}$ ($i > j$). The upper-right triangular region of $\mathcal{P}$ are padded with zeros.

The prefix structure imposes the following design constraint: each prefix node must have exactly one pair of upper/lower parent:
\begin{equation}
    \mathcal{P}[i,j] = 1 \implies 
    \exists \: k \in [j+1, i] \cap \mathbb{Z}, \quad \mathcal{P}[i,k] = \mathcal{P}[k-1,j] = 1
\label{eq:prefix_constraint}
\end{equation}
When multiple $k$ satisfy Equation~\ref{eq:prefix_constraint}, the minimal $k$ determines the canonical parent pair. For example, node $(5,0)$ in Fig.~\ref{fig:prefix_representation} have two valid parent pairs $(5,5)/(4,0)$ and $(5,4)/(3,0)$, with the latter selected as the parent nodes. Consequently, each bitmap $\mathcal{P}$ uniquely defines a valid parallel prefix tree.

\begin{figure}[!t]
  \centering
  \includegraphics[width=0.9\linewidth]{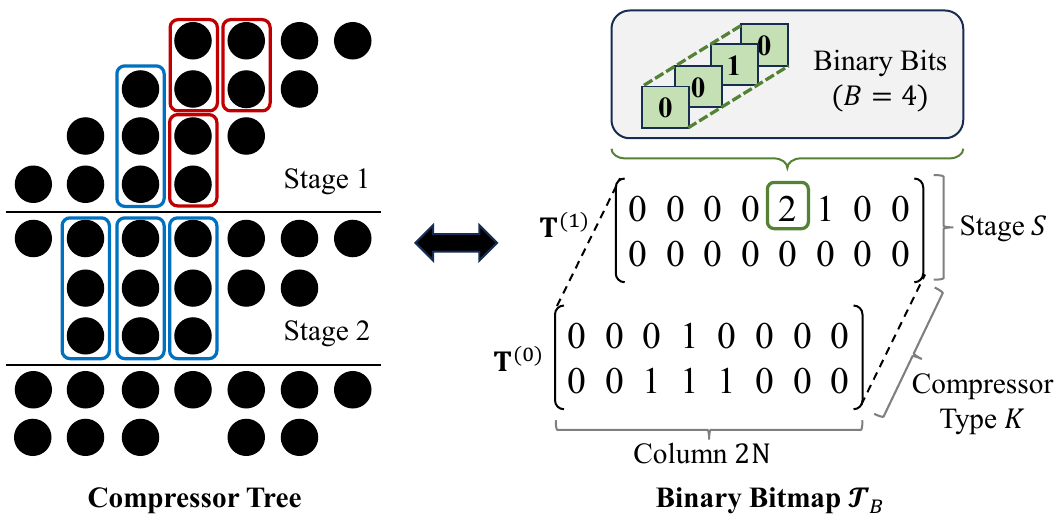}
  \caption{Data representation of compressor tree.}
  \label{fig:ct_representation}
\end{figure}

\noindent \textbf{Compressor Tree Representation.}
As illustrated in Fig.\ref{fig:ct_representation}, we represent a compressor tree using a three-dimensional tensor $\mathcal{T} \in \mathbb{R}^{K \times 2N \times S}$, where $K$ denotes the number of compressor types, $N$ denotes the multiplier bit-width, and $S$ denotes the number of stages.
Without loss of generality, we consider $K = 2$ and only employ full adders and half adders to construct CT in \papername{}.
The maximum compression stage $S$ is computed as $S_{min} + \Delta S$. $S_{min}$ is the stage count of an $N$-bit Wallace multiplier, which is also the theoretical lower bound~\cite{wallace1964suggestion}. $\Delta S = 1$ is a hyperparameter that balances design quality against the number of feasible solutions.

To formalize the CT design rules, we introduce tensor $\mathcal{V} \in \mathbb{R}^{2N \times (S+1)}$ that tracks the number of partial products at each compression stage, with $\mathcal{V}[\cdot, 0]$ determined by PPG and $\mathcal{V}[\cdot, S]$ as the final partial product counts. Using the established notations, a valid compressor tree design $\mathcal{T}$ should satisfy the following constraints:
\begin{equation}
    \mathcal{T}[k, c, s] \geq 0, \quad \forall k\in \{0, 1\}, c \in [0, 2N), s \in [0, S)
\label{eq:ct_drc_nonzero}
\end{equation}
\begin{equation}
    3 \cdot \mathcal{T}[0, c, s] + 2 \cdot \mathcal{T}[1, c, s] \leq \mathcal{V}[c, s], \quad \forall c \in [0, 2N), s \in [0, S)
    \label{eq:ct_drc_pp}
\end{equation}
\begin{equation}
\begin{aligned}
    & \mathcal{V}[c, s+1] = \mathcal{V}[c, s] - (2 \cdot \mathcal{T}[0, c, s] + \mathcal{T}[1, c, s]) \\
    & + (\mathcal{T}[0, c-1, s] + \mathcal{T}[1, c-1, s]), ~\forall c \in [1, 2N), s \in [0, S)
    \label{eq:ct_drc_next}
\end{aligned}
\end{equation}
\begin{equation}
    \mathcal{V}[0, s+1] = \mathcal{V}[0, s] - (2 \cdot \mathcal{T}[0, 0, s] + \mathcal{T}[1, 0, s]), \quad \forall s \in [0, S)
    \label{eq:ct_drc_next0}
\end{equation}
\begin{equation}
    0 \leq \mathcal{V}[c, S] \leq 2, \quad \forall c \in [0, 2N)
    \label{eq:ct_drc_final}
\end{equation}

Given a valid CT configuration $\mathcal{T}$, the mapping from $\mathcal{T}$ to RTL implementation is not unique, as input assignments of co-located compressors can be permuted without affecting the multiplication correctness.
To obtain a unique and high-quality CT design, we adopt a deterministic interconnection scheme that prioritizes assigning input bits with minimal signal arrival time to FAs over HAs. The internal delay of compressors is estimated on the selected technology node.

\noindent \textbf{Training Diffusion Model.}
The training process of diffusion models proceeds as follows:
\begin{enumerate}
    \item We convert the circuit representations into binary bitmaps. For CT designs, we follow Analog-Bit~\cite{chen2022analog} to split integers into multiple binary bits with different weights. The resulting bitmap is denoted as $\mathcal{T}_B$, as shown in Fig.~\ref{fig:ct_representation}.
    \item We cast the input bitmap to a real-valued tensor, denoted as $x_0$. Each binary bit $b \in \{0, 1\}$ is mapped to a corresponding real value $r \in \{-1.0, 1.0\}$. Conversely, a real-valued $r$ can be quantized back to a discrete $b$ according to each element's sign.
    \item We randomly sample Gaussian noise $\epsilon \sim \mathcal{N}(0, \mathbf{I})$ and obtain noisy data $x_t$ according to the noise schedule in Equation~\eqref{eq:diffusion_forward}.
    \item We train the diffusion model $\epsilon_\theta$ to predict the injected noise as in Equation~\eqref{eq:diffusion_reverse}. $\epsilon_\theta$ is implemented with U-Net backbones~\cite{ronneberger2015u}. The training process can be formulated as minimizing the mean-square error (MSE) loss:
    \begin{equation}
        \min_\theta \mathcal{L}_\theta(x_t, \epsilon, t) = (\epsilon_\theta(x_t, t) - \epsilon)^2.
    \end{equation}
\end{enumerate}


\noindent \textbf{Training Cost Predictor.}
Apart from training diffusion models, \papername{} also utilizes neural cost predictors to guide the circuit generation.
Specifically, we define function $f$ as the mapping from an arithmetic circuit design to the post-synthesis Quality-of-Results (QoR).
To enable gradient-based guidance, $f$ takes in the same continuous tensor $x$ as the diffusion model $\epsilon_\theta$. 
To trade off multiple competing QoR metrics and obtain Pareto-optimal circuit design, we conduct multiple synthesis flows to evaluate the arithmetic circuit design $x$. The considered design scenarios include timing-driven, area-driven, and balanced optimization. The total cost $f(x)$ is defined as a weighted sum of delay and area across all the scenarios:
\begin{equation}
    f(x) = (w \sum_{i=1}^{n} \text{delay}_i(x) + (1-w) \sum_{i=1}^{n} \text{area}_i(x)) / n.
    \label{eq:cost_function}
\end{equation}
where $\text{delay}_i$ and $\text{area}_i$ are the post-synthesis metrics in $i$-th design scenario, and $w$ is the weight to trade off the competing goals.

Since $f$ lacks a closed-form expression, we aim to train a differentiable regression model $f_\pi$ to approximate the ground truth $f$. 
We implement $f_\pi$ as a 3-layer CNN composed of convolutional residual blocks~\cite{he2016deep}.
Given design $x$ labeled with its corresponding QoR $y$, the training process is formulated as minimizing the MSE loss:
\begin{equation}
    \min_\pi \mathcal{L}_\pi(x, y) = (f_\pi(x)-y)^2
\end{equation}

\subsection{Conditional Design Generation}
\label{sec:sampling}

\begin{figure}[!t]
  \centering
  \includegraphics[width=\linewidth]{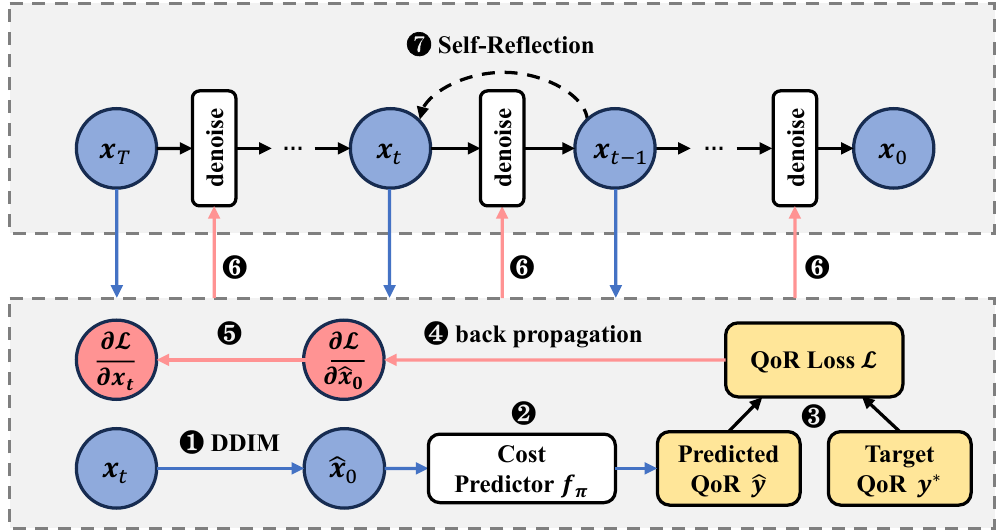}
  \caption{Conditional generation of arithmetic circuits. Gradient-guided sampling (Step 1-6) drives the generated designs towards achieving the target quality-of-results (QoR). Self-reflection (Step 7) prevents the gradient guidance from compromising the structural correctness.}
  \label{fig:cond_synth}
\end{figure}

To consistently explore high-potential designs, \papername{} introduces a conditional sampling process.
Conceptually, we incorporate the gradient guidance from the neural cost predictor $f_\pi$ to steer the denoising direction towards high QoR outcomes.
As illustrated in Fig.~\ref{fig:cond_synth}, the conditional design generation process comprises two key ingredients: gradient-guided sampling and self-reflection.

\noindent \textbf{Gradient-Guided Sampling.} 
A critical challenge in guiding the diffusion process is evaluating the quality of intermediate states. On the one hand, the noisy $x_t$ lacks a well-defined structure. Since the cost predictor $f_\pi$ is trained on unperturbed tensors, directly assessing $x_t$ cannot reliably provide meaningful guidance.
On the other hand, the desired $x_0$ is available upon completion of the diffusion process.

To address this dilemma, we draw inspiration from DDIM~\cite{song2020denoising} to approximate $x_0$ at intermediate steps (Step \blackcircle{1}):
\begin{equation}
    \hat{x}_0 = \dfrac{x_t - \sqrt{1 - \alpha_t} \cdot \epsilon_\theta(x_t, t)}{\sqrt{\alpha_t}},
    \label{eq:x0_prediction}
\end{equation}
DDIM originally employs the predicted clean data $\hat{x}_0$ to accelerate sampling, whereas \papername{} repurposes this mechanism to acquire informative feedback from cost predictor $f_\pi$. Although $\hat{x}_0$ remains an imperfect approximation, it exhibits a more regularized structure over the noisy $x_t$, which enables cost predictor $f_\pi$ to deliver a more accurate prediction on post-synthesis QoR (Step \blackcircle{2}).

For efficient design optimization, we expect to prune suboptimal candidate designs whose predicted QoR $\Tilde{y}=f_\pi(\hat{x}_0)$ does not reach a desirable target $y^*$.
This objective is formulated as minimizing loss $\mathcal{L}(y^*, \Tilde{y})$, where we select mean-square error as the loss function (Step \blackcircle{3}).
To achieve effective minimization, we backpropagate the gradient of loss $\mathcal{L}$ down to $x_t$ (Steps \blackcircle{4}-\blackcircle{5}), and add a gradient rectification term $\frac{\partial \mathcal{L}}{\partial x_t}$ in the denoising step from $x_t$ to $x_{t-1}$ (Step \blackcircle{6}):
\begin{equation}
    x_{t-1} = \mathcal{S}(x_t, \hat{\epsilon}, t) - s(t) \cdot \nabla_{x_t} \mathcal{L}(y^*, f_\pi(\hat{x}_0)),
    \label{eq:guided_sampling}
\end{equation}
This adjustment attempts to progressively reduce $\mathcal{L}$ throughout the denoising process. In Equation~\eqref{eq:guided_sampling}, $s(t) > 0$ controls the guidance strength, and $\mathcal{S}$ adopts the sampling procedure from DDIM~\cite{song2020denoising}.

\noindent \textbf{Self-Reflection.} 
In practice, the gradient-guided diffusion process struggles to balance design correctness and quality.
A radical guidance strength $s(t)$ may sabotage the normal denoising process and lead to invalid designs. Whereas a conservative one may not effectively minimize loss $\mathcal{L}(y, y^*)$ in limited denoising steps.
To address this issue, we apply the self-reflection strategy following~\cite{bansal2023universal}, which ensures QoR optimization without compromising correctness.
Specifically, after gradient-guided denoising produces $x_{t-1}$, \papername{} reintroduces random Gaussian noise $\epsilon'$ back to $x_{t-1}$ (Step \blackcircle{7}):
\begin{equation}
    x_t = \sqrt{\alpha_t / \alpha_{t-1}} \cdot x_{t-1} + \sqrt{1 - \alpha_t / \alpha_{t-1}} \cdot \epsilon',
    \label{eq:reinject_noise}
\end{equation}
The resulting intermediate state $x_t$ has an in-distribution pattern at timestep $t$ and better alignment with the gradient guidance. In plain words, self-reflection mechanism enables selective sampling of $x_t$ instances that largely satisfy the gradient guidance.
This iterative process is repeated multiple times before advancing to timestep $t-2$.

Algorithm~\ref{algo:guided_sampling} summarizes the overall conditional design generation process, which integrates the proposed gradient-guided sampling and self-reflection mechanism.
While it largely respects the design constraints, it cannot perfectly guarantee structural integrity, which necessitates the legalization process introduced in Section~\ref{sec:legalization}.

\begin{algorithm}[!t]
\caption{Gradient-Guided Sampling}
\begin{algorithmic}[1]
\Statex \textbf{Input:} noisy data $x_T \sim \mathcal{N}(0, I)$, total timestep $T$, noise scale $\{\alpha_t\}_{t=1}^{T}$, diffusion model $\epsilon_\theta$, cost predictor $f_\pi$, criterion $\mathcal{L}$, target QoR $y^*$, guidance strength $s(t)$, and self-reflection step $k$.
\Statex \textbf{Output:} Sampled design $x_0$
\For{$t = T, T-1, \dots, 1$}
    \For{$n = 1, 2, \dots, k$}
        \State Predict noise $\hat{\epsilon} = \epsilon_\theta(x_t, t)$
        \State Predict final design $\hat{x}_0$ as Equation~\eqref{eq:x0_prediction}
        \State Denoise design $x_t$ to $x_{t-1}$ as Equation~\eqref{eq:guided_sampling}
        \State Perturb $x_{t-1}$ to $x_t$ as Equation~\eqref{eq:reinject_noise}
    \EndFor
\EndFor
\State \Return $x_0$
\end{algorithmic}
\label{algo:guided_sampling}
\end{algorithm}

\begin{algorithm}[!t]
\caption{Compressor Tree Legalization}
\begin{algorithmic}[1]
\Statex \textbf{Input:} Compressor tree $\mathcal{T}$ to be legalized
\Statex \textbf{Output:} Legalized compressor tree structure
\While{$\exists$ error $e \in E(\mathcal{T})$}
    \State Initialize candidate action set $\mathcal{A} \leftarrow \emptyset$
    \If{$e$ is over-compression at column $c$ and stage $s$} \label{line:over_compression_start}
        \If{$c > 1$}
            \State $\mathcal{A}  \leftarrow \mathcal{A}  \cup \{\texttt{SplitFA}_{(c-1), s'} | 1 \leq s' < s \}$
        \EndIf
        \State $\mathcal{A}  \leftarrow \mathcal{A}  \cup \{\texttt{ReplaceFA}_{c, s'} | 1 \leq s' \leq s \}$
        \State $\mathcal{A}  \leftarrow \mathcal{A}  \cup \{\texttt{DeleteHA}_{c, s}\}$ \label{line:over_compression_end}
    \ElsIf{$e$ is under-compression at column $c$} \label{line:under_compression_start}
        \If{$c > 1$}
            \State $\mathcal{A} \leftarrow \mathcal{A} \cup \{\texttt{FuseFA}_{(c-1), s'} | 1 \leq s' < S \}$
        \EndIf
        \State $\mathcal{A} \leftarrow \mathcal{A} \cup \{\texttt{ReplaceHA}_{c, s'} | 1 \leq s' \leq S \}$
        \State $\mathcal{A} \leftarrow \mathcal{A} \cup \{\texttt{AddHA}_{c, s'} | 1 \leq s' \leq S \}$ \label{line:under_compression_end}
    \EndIf
    \State Choose action $A^* = \text{argmin}_{A \in \mathcal{A}} |E(A(\mathcal{T}))|$ \label{line:choose_action}
    \State Update design $\mathcal{T} \leftarrow A^*(\mathcal{T})$ \label{line:apply_action}
\EndWhile
\State \Return $\mathcal{T}$
\end{algorithmic}
\label{algo:legalization}
\end{algorithm}

\subsection{Design Legalization}
\label{sec:legalization}

To ensure the functional correctness of the generated arithmetic circuit designs, \papername{} introduces heuristic legalization procedures to eliminate design rule violations (DRVs) in the circuits' tensor representation. 
For prefix adders, we adopt the established approach from PrefixRL~\cite{roy2021prefixrl}, which iterates through all the prefix nodes and fills the missing lower parents to ensure a valid bitmap $\mathcal{P}$.
For compressor trees, GOMIL~\cite{xiao2021gomil} utilizes integer linear programming (ILP) to produce valid designs satisfying Equation~\eqref{eq:ct_drc_nonzero}-\eqref{eq:ct_drc_final}. However, the unpredictable behavior of ILP solver may negate the benefits from conditional circuit generation, rendering it unsuitable in our use case. 

Empirically, we observe that CT designs generated by the diffusion models typically exhibit few DRVs, which provide opportunities to correct these errors via efficient heuristics.
As shown in Algorithm~\ref{algo:legalization}, \papername{} adopts CT legalization procedure that takes a series of local adjustments to resolve design rule violations.
We denote $E(\mathcal{T})$ as the set of error positions in the compressor tree design $\mathcal{T}$, consisting of over-compression (violating Equation~\eqref{eq:ct_drc_pp}) and under-compression (violating Equation~\eqref{eq:ct_drc_final}).
The algorithm handles over-compression by increasing partial products or reducing compressors (Lines \ref{line:over_compression_start}-\ref{line:over_compression_end}), and addresses under-compression by adding compressors if needed (Lines \ref{line:under_compression_start}-\ref{line:under_compression_end}).
For quick convergence to a valid configuration, the algorithm greedily selects an action $A^*$ from all candidate actions $\mathcal{A}$, which minimizes the size of $E(\mathcal{T})$ (Line \ref{line:choose_action} to \ref{line:apply_action}).
When no action can effectively shrink $E(\mathcal{T})$, we force the legalization process to take a detour, selecting an action that incurs the minimum new errors.
In most cases, we observe that Algorithm~\ref{algo:legalization} manages to quickly converge on valid solutions.

\subsection{Model Fine-Tuning}
\label{sec:finetuning}

To further improve optimization efficiency, \papername{} fine-tunes the models leveraging the explored high-quality designs.
This process enables the models to better capture the intricate pattern of optimized designs, which increases the likelihood of discovering even superior ones.
As shown in Fig.~\ref{fig:framework_overview}, \papername{} performs $M$ rounds of optimization.
In each round, the generated designs constitute a new unlabeled dataset $\mathcal{D}'$.
A subset of $\mathcal{D}'$ is evaluated for QoR with physical synthesis tools, creating a labeled dataset $\hat{\mathcal{D}}'$.
The newly acquired datasets are then merged with the initial dataset $\mathcal{D}$ and $\hat{\mathcal{D}}$ to fine-tune the diffusion model $\epsilon_\theta$ and regression model $f_\pi$.


\section{Evaluations}
\label{sec:experiment}

\subsection{Setup}
\label{sec:experiment_setup}

\noindent \textbf{Platform.}
All experiments are conducted on a Linux-based platform equipped with a 2.9GHz 64-core Intel Xeon Gold 6226R CPU, 376GB of memory, and an NVIDIA A6000 GPU.
We use open-sourced logic synthesis tool Yosys (version 0.27)~\cite{Yosys} and the physical synthesis tool OpenROAD (version 2.0)~\cite{ajayi2019openroad} to assess the Quality-of-Results (QoR) of multiplier designs using the NanGate 45nm technology process library~\cite{nangate2008freepdk45}.

\noindent \textbf{Dataset Preparation.}
Training robust generative models requires an extensive dataset of validated design configurations. 
To meet this demand, we utilize heuristic algorithms to modify manual designs~\cite{sklansky1960conditional,kogge1973parallel,brent1982regular,wallace1964suggestion,dadda1990some} for improved theoretical properties. The modification operation and associated legalization process are adopted from RL-MUL~\cite{zuo2023rl} and PrefixRL~\cite{moto2018prefix} for CT and CPA, respectively. For CT designs, we add an extra optimization by randomly swapping compressors at different stages, yielding new CTs beyond the design space of RL-MUL.

\noindent \textbf{Design Verification.}
For RTL generation of \papername{}, we implement a Python program to convert the compressor tree configurations $\mathcal{T}$ and prefix adder configurations $\mathcal{P}$ into Verilog HDL codes. These components are then composed to construct multipliers.
The pre-synthesis Verilog codes of all multipliers, including those from baseline methods and from \papername{}, are converted to and-inverter graph with ABC~\cite{brayton2010abc} and formally verified with RevSCA-2.0~\cite{mahzoon2021revsca}. 

\noindent \textbf{Baselines.}
We compare \papername{} with four baseline methods for multiplier optimization, including Wallace multiplier~\cite{wallace1964suggestion}, ILP-based method GOMIL~\cite{xiao2021gomil}, and RL-based methods RL-MUL~\cite{zuo2023rl} and ArithmeticTree~\cite{lai2024scalable}.

\noindent \textbf{Hyperparameter Settings.}
We implement the diffusion models $\epsilon_\theta$ and cost predictors $f_\pi$ with PyTorch.
To evaluate the QoR cost function as in Equation~\eqref{eq:cost_function}, we synthesize multiplier designs under area-driven scenario and timing-driven scenario, and normalize the delay and area with corresponding values of Wallace multiplier~\cite{wallace1964suggestion}.
The trade-off weight is set to $w = 0.66$.
For gradient-guided sampling, we set the target QoR objective $y=0.7$, the guidance strength $s(t) = 10 \sqrt{1 - \alpha_t}$, and the self-reflection step to $k=25$.
We conduct a comprehensive optimization of multipliers covering both CT and CPA.
We commence by optimizing CT and fix CPA with the default adder from the synthesis tool.
Then we adopt the best explored CT and refine the CPA.
For CT and CPA, respectively, the initial dataset consists of $|D_0| \approx 15000$ unlabeled samples, and $|\hat{D}_0| = 1000$ designs labeled with post-synthesis QoR.
Optimization runs for $M=5$ rounds, where in each round we sample $|D_m| = 1000$ designs and randomly select $|\hat{D}_m| = 100$ designs for physical synthesis.
In total, the synthesis tools are invoked 3000 times, which is the same as learning-based baselines RL-MUL and ArithmeticTree.

\begin{figure}[!t]
  \centering
  \includegraphics[width=\linewidth]{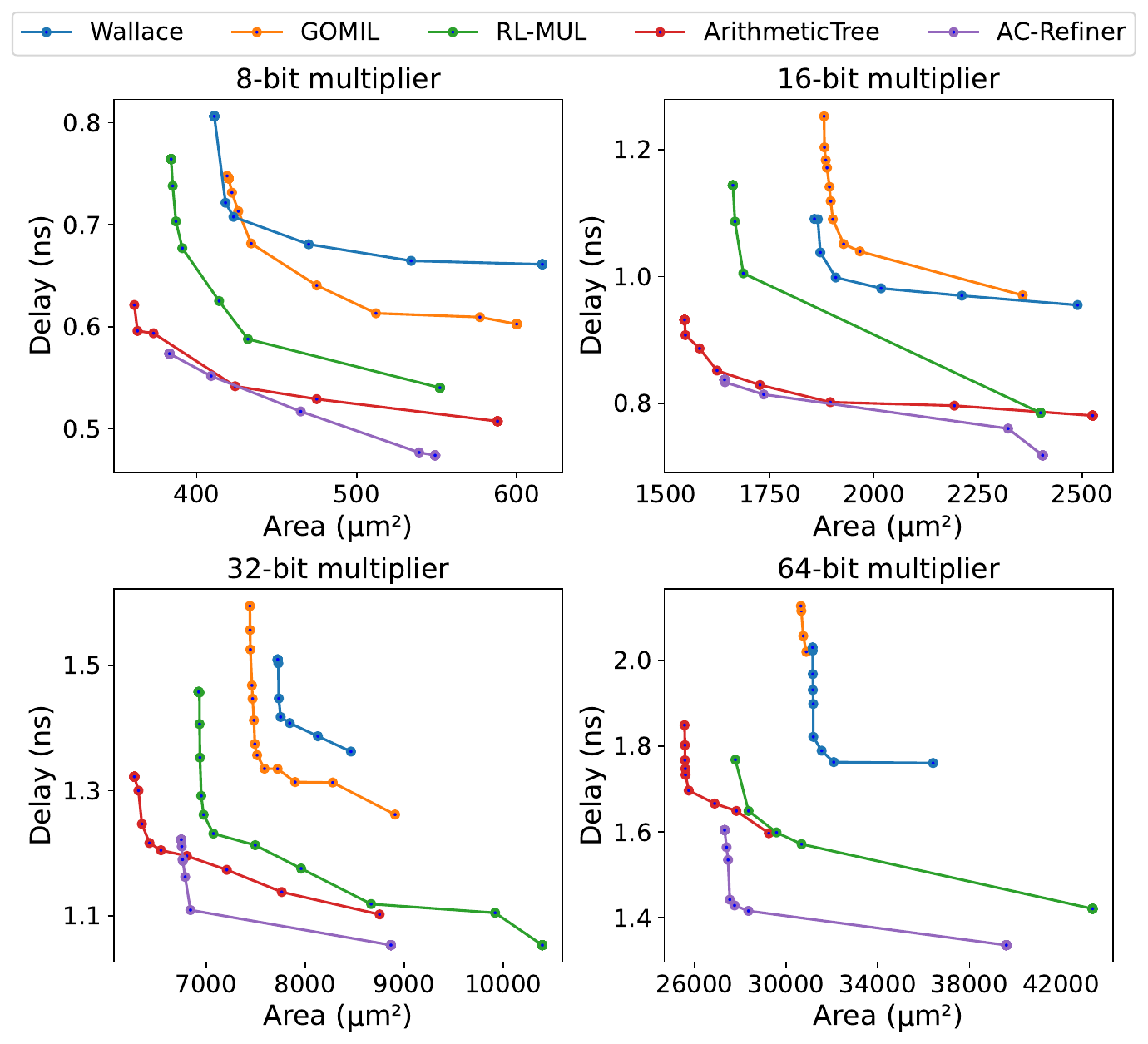}
  \caption{Pareto frontiers of the synthesized results on multipliers.}
  \label{fig:sweep_delay}
\end{figure}

\subsection{Multiplier Optimization Result}
\label{sec:compare_multiplier}

We compare the Pareto frontiers obtained using different optimization algorithms. 
Following established evaluation methodologies~\cite{zuo2023rl, lai2024scalable}, we select the optimal multiplier from each approach and synthesize them with a target delay ranging from 0ns to 2ns to demonstrate the trade-off between timing and area.
As shown in Fig.~\ref{fig:sweep_delay}, \papername{} achieves superior Pareto optimality over all baselines.
Both the manually designed Wallace multiplier and the algorithmic method GOMIL fail to account for the complexities inherent in physical implementation, leading to suboptimal post-synthesis delay and area.
RL-MUL modifies only the column-wise compressor count without considering detailed stage-specific compressor assignment or CPA optimization, yielding inferior performance.
ArithmeticTree trains the RL agents to construct multipliers in a gate-by-gate manner. Despite fine-grained optimization, it suffers from exploration inefficiency as the RL agents must rely on extensive trial-and-error iterations. 
Compared with the state-of-the-art ArithmeticTree method, \papername{} achieves a delay reduction of up to 15\% and an area reduction of up to 10\%.

\begin{figure}[!t]
  \centering
  \includegraphics[width=0.9\linewidth]{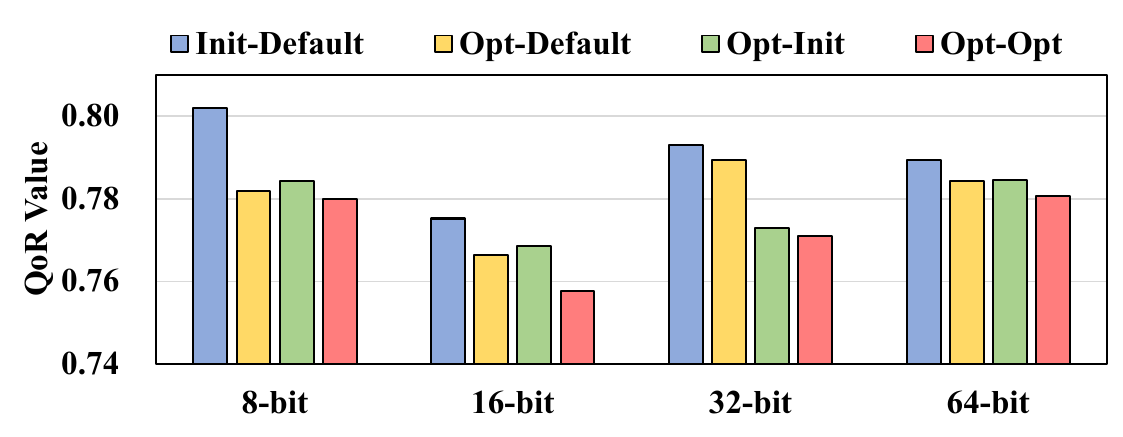}
  \caption{Effect on QoR value upon progressive inclusion of diffusion-driven optimization. Labels indicate the source of CT and CPA.}
  \label{fig:progressive_optimization}
\end{figure}

\begin{table}[!t]
\centering
\caption{Systolic array comparison on delay, area, and area-delay product (ADP).}
\resizebox{\linewidth}{!}{%
\begin{tabular}{P{4.2em}|P{8.3em}|P{2em}P{2em}P{3.3em}|P{2em}P{2em}P{3.3em}}
\toprule
\multirow{2}{*}{Objective} & \multirow{2}{*}{Method} & \multicolumn{3}{c|}{8-bit} & \multicolumn{3}{c}{16-bit} \\ \cline{3-8}
 & & \thead{Delay\\(ns)} & \thead{Area\\($\mu\text{m}^2$)} & \thead{ADP\\($\mu\text{m}^2\cdot$ns)} & \thead{Delay\\(ns)} & \thead{Area\\($\mu\text{m}^2$)} & \thead{ADP\\($\mu\text{m}^2\cdot$ns)} \\ \midrule
\multirow{4}{*}{Min Delay} 
& Wallace \cite{wallace1964suggestion}            & 1.0489 & 18896 & 19820 & 1.4164 & 60672 & 85935 \\
& GOMIL \cite{xiao2021gomil}                      & 1.1101 & 20384 & 22628 & 1.5559 & 58864 & 91586 \\
& RL-MUL \cite{zuo2023rl}                         & 0.9746 & 18592 & 18119 & 1.3502 & 56176 & 75848 \\
& ArithmeticTree \cite{lai2024scalable}           & 0.9351 & 19440 & 18178 & 1.3311 & 60832 & 80973 \\
& \textbf{\papername{} (Ours)}                    & \textbf{0.9317} & 19376 & \textbf{18052} & \textbf{1.2723} & \textbf{55488} & \textbf{70597} \\ \midrule
\multirow{4}{*}{Trade-off}   
& Wallace \cite{wallace1964suggestion}            & 1.2846 & 14816 & 19032 & 1.7833 & 45056 & 80348 \\
& GOMIL \cite{xiao2021gomil}                      & 1.2661 & 15056 & 19062 & 1.8992 & 45520 & 86451 \\
& RL-MUL \cite{zuo2023rl}                         & 1.1825 & 14496 & 17141 & 1.6443 & 42048 & 69139 \\
& ArithmeticTree \cite{lai2024scalable}           & 1.1163 & 14144 & 15788 & 1.5504 & 40544 & 62859 \\
& \textbf{\papername{} (Ours)}                    & \textbf{1.0479} & 14432 & \textbf{15123} & \textbf{1.3803} & 41712 & \textbf{57575} \\ \bottomrule
\end{tabular}
}%
\label{tab:mult_pe_result}
\end{table}

\begin{figure}[t]
    \centering
    
    \begin{minipage}{0.48\columnwidth}
        \centering
        \includegraphics[width=\linewidth]{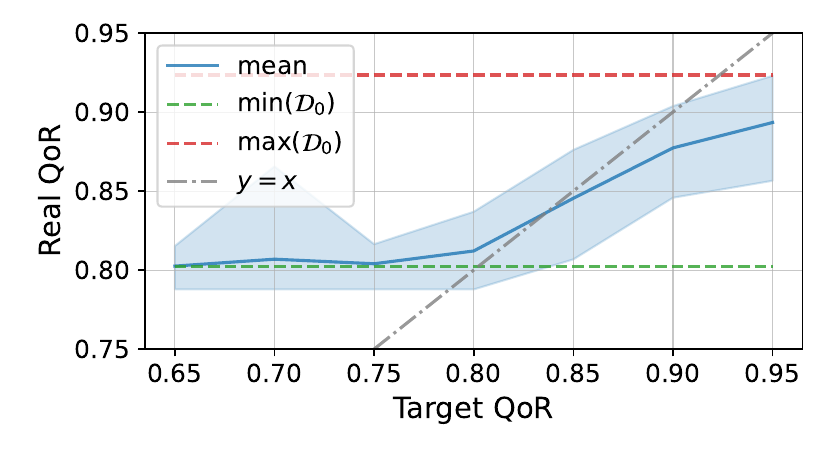}
        \parbox{\linewidth}{\centering (a) 8-bit compressor tree}
    \end{minipage}
    \hspace{2pt}
    \begin{minipage}{0.48\columnwidth}
        \centering
        \includegraphics[width=\linewidth]{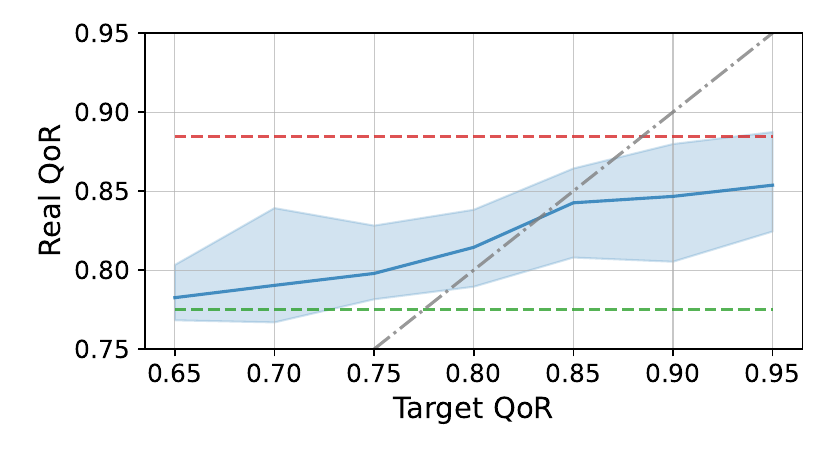}
        \parbox{\linewidth}{\centering (b) 16-bit compressor tree}
    \end{minipage}
    
    \begin{minipage}{0.48\columnwidth}
        \centering
        \includegraphics[width=\linewidth]{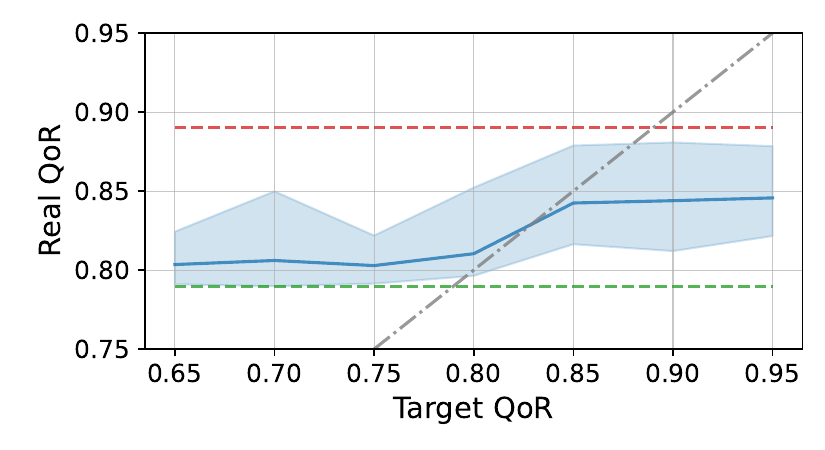}
        \parbox{\linewidth}{\centering (c) 32-bit compressor tree}
    \end{minipage}
    \hspace{2pt}
    \begin{minipage}{0.48\columnwidth}
        \centering
        \includegraphics[width=\linewidth]{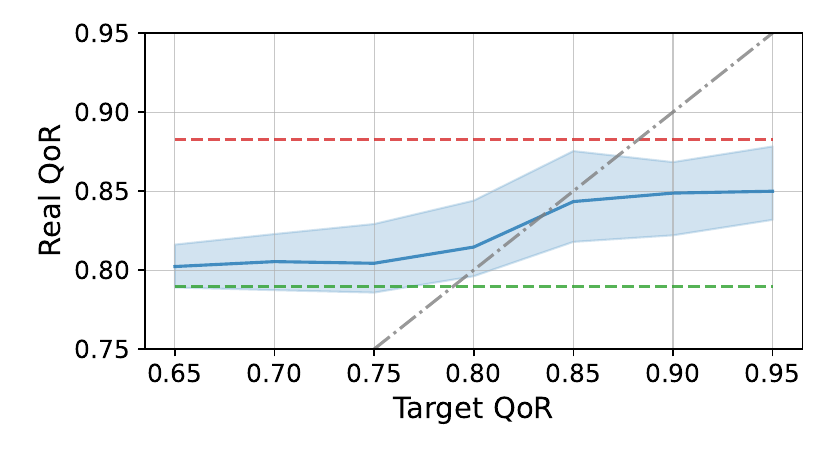}
        \parbox{\linewidth}{\centering (d) 64-bit compressor tree}
    \end{minipage}
    
    \caption{Plot of real QoR versus target QoR for guided sampling. Dotted lines represent the min/max QoR of initial dataset $\hat{D_0}$ and the ideal line $y = x$. The boundary of shaded regions indicates the min/max QoR of sampled designs.}
    \label{fig:impact_cond}
\end{figure}

\begin{figure}[t]
    \centering
    
    \begin{minipage}{0.48\columnwidth}
        \centering
        \includegraphics[width=\linewidth]{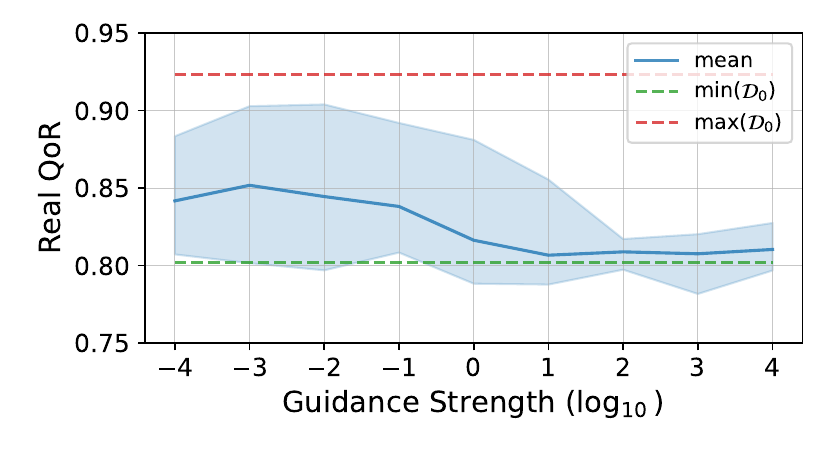} \\
        \parbox{\linewidth}{\centering (a) 8-bit compressor tree}
    \end{minipage}
    \hspace{2pt}
    \begin{minipage}{0.48\columnwidth}
        \centering
        \includegraphics[width=\linewidth]{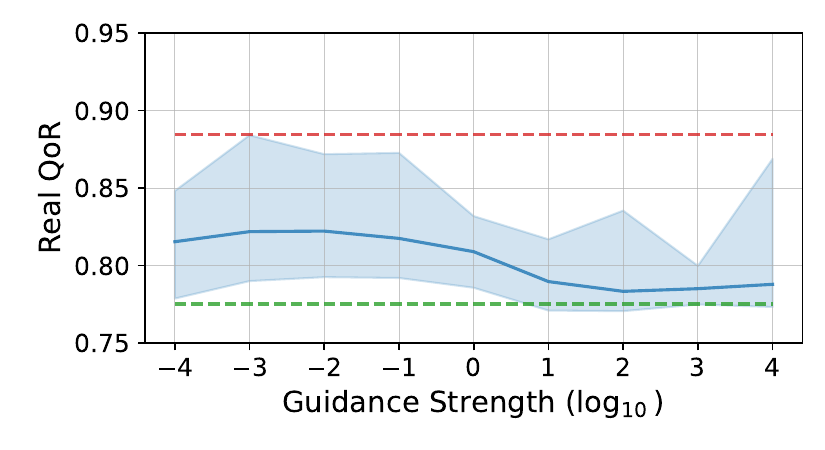} \\
        \parbox{\linewidth}{\centering (b) 16-bit compressor tree}
    \end{minipage}
    
    \begin{minipage}{0.48\columnwidth}
        \centering
        \includegraphics[width=\linewidth]{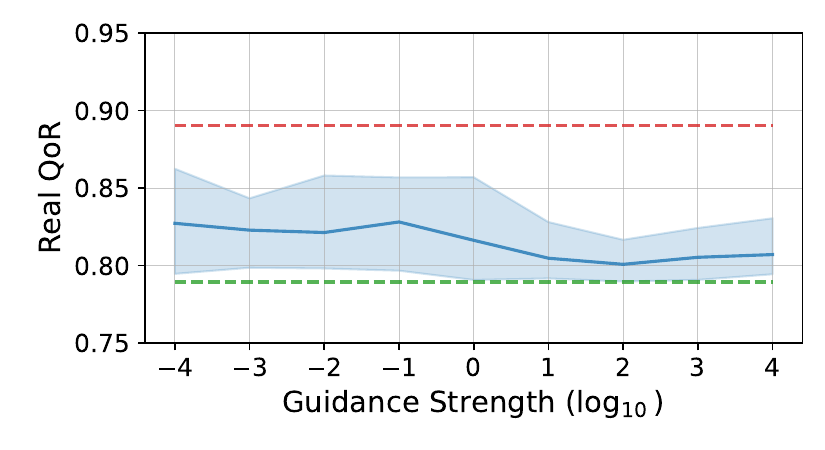} \\
        \parbox{\linewidth}{\centering (c) 32-bit compressor tree}
    \end{minipage}
    \hspace{2pt}
    \begin{minipage}{0.48\columnwidth}
        \centering
        \includegraphics[width=\linewidth]{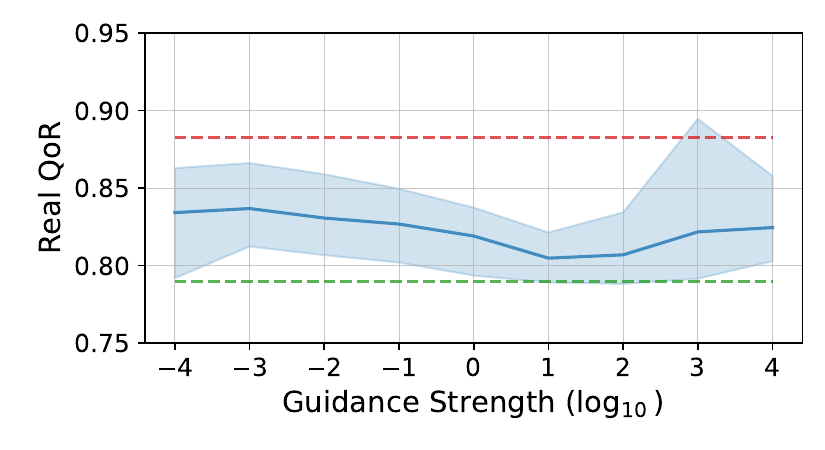} \\
        \parbox{\linewidth}{\centering (d) 64-bit compressor tree}
    \end{minipage}
    
    \caption{Plot of real QoR versus guidance strength for guided sampling. Dotted lines represent the min/max QoR of initial dataset $\hat{D_0}$. Shaded region boundaries indicates min/max QoR of sampled designs.}
    \label{fig:impact_strength}
\end{figure}

We also examine the QoR improvement by progressively incorporating diffusion-driven optimization.
Fig.~\ref{fig:progressive_optimization} shows that CTs and CPAs optimized by \papername{} (Opt) achieve better QoR over the ones in the initial dataset (Init).
Notably, the initial CPAs may not integrate well with the optimized CT than the default adders from the synthesis tool (Default), yet \papername{} still identifies CPAs that enhance the overall QoR of multipliers.

To demonstrate the effectiveness of \papername{} in real-world designs, we incorporate multipliers from all approaches into multiply-accumulators of systolic arrays, which are commonly adopted in AI accelerators~\cite{jouppi2017datacenter}.
As shown in TABLE~\ref{tab:mult_pe_result}, \papername{} retains its performance advantages across different scenarios, achieving the lowest delay when optimizing for timing, and the smallest area-delay product (ADP) when balancing timing and area.

\subsection{Ablations}
\label{sec:ablation}
We perform ablation studies on the components of \papername{} to study their individual effects: gradient-guided sampling, self-reflection, and model fine-tuning.

\noindent \textbf{Gradient-Guided Sampling.} 
The QoR outcome $\tilde{y}$ of the conditional sampling process is influenced by two factors: target QoR $y^*$ and guidance strength $s(t)$. We demonstrate their individual effects through the optimization of compressor trees across various bit-width. 

In Fig.~\ref{fig:impact_cond} we examine the impact of $y^*$ on $\tilde{y}$.
When $y^*$ falls in the range of the initial dataset, the real QoR $\tilde{y}$ shows a strong correlation with $y^*$. This phenomenon shows that the neural cost predictor successfully captures the nuanced interplay between circuit structure and physical synthesis, offering effective guidance for conditional circuit generation.
When $y^*$ is set extremely low, the real QoR outcome generalizes beyond the minima of the initial dataset, highlighting the efficacy of diffusion-driven optimization in expanding Pareto frontiers.

Fig.~\ref{fig:impact_strength} shows the effect of $s(t)$ on $\tilde{y}$, where the x-axis represents the logarithmic values of the coefficient in $s(t)$.
Weak guidance strength leads to QoR distributions resembling the initial dataset. Whereas overly strong guidance disrupts the diffusion process, resulting in degraded QoR.
Therefore, proper tuning of $s(t)$ is critical for achieving effective optimization outcomes.


\begin{table}[!t]
\centering
\caption{Results of average steps to legalize sampled compressor trees. The maximum legalization step is set to 5000.}
\begin{tabular}{ >{\centering\arraybackslash}p{2.8cm} | >{\centering\arraybackslash}p{1cm} | >{\centering\arraybackslash}p{1cm} | >{\centering\arraybackslash}p{1cm} | >{\centering\arraybackslash}p{1cm}}
\toprule
Method             & 8-bit & 16-bit & 32-bit & 64-bit \\ \midrule
Unconditional      & 0     & 4      & 33     & 103   \\
Conditional w/o SR & 7     & 4      & 179    & 81   \\
Conditional w/  SR & 0     & 2      & 1      & 7    \\ \midrule
VAE~\cite{kingma2013auto} & 26    & 3757   & N/A   & N/A  \\
\bottomrule
\end{tabular}
\label{tab:legalization}
\end{table}

\begin{figure}[t]
    \centering
    
    \begin{minipage}{0.45\columnwidth}
        \centering
        \includegraphics[width=\linewidth]{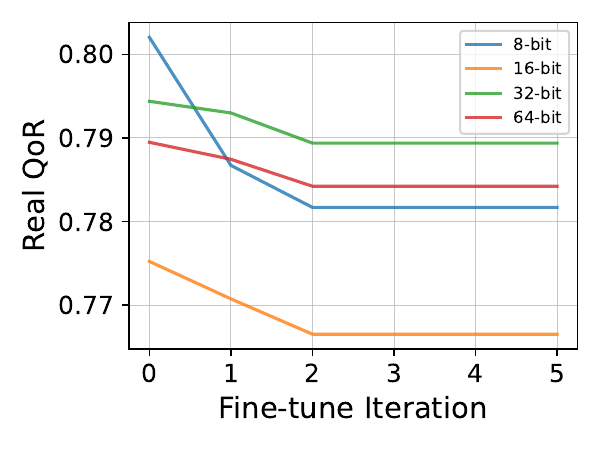} \\
        \parbox{\linewidth}{\centering (a) Compressor tree}
    \end{minipage}
    \hspace{0pt}
    \begin{minipage}{0.45\columnwidth}
        \centering
        \includegraphics[width=\linewidth]{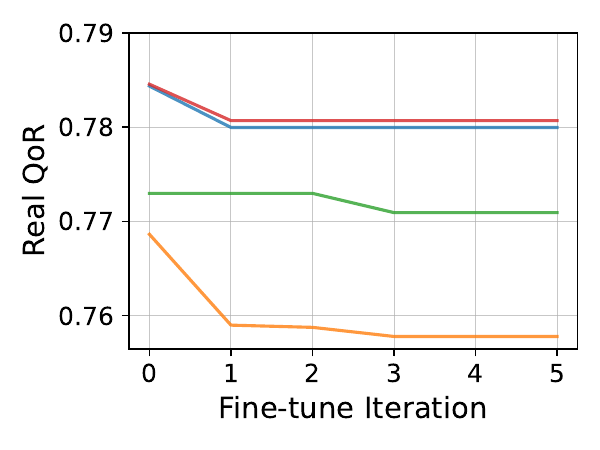} \\
        \parbox{\linewidth}{\centering (b) Prefix tree}
    \end{minipage}
    
    \caption{The effect of fine-tuning iteration on QoR optimization results. 0-th iteration represents minimum QoR from initial labeled dataset.}
    \label{fig:impact_finetune}
\end{figure}

\begin{figure}[!t]
  \centering
  \includegraphics[width=\linewidth]{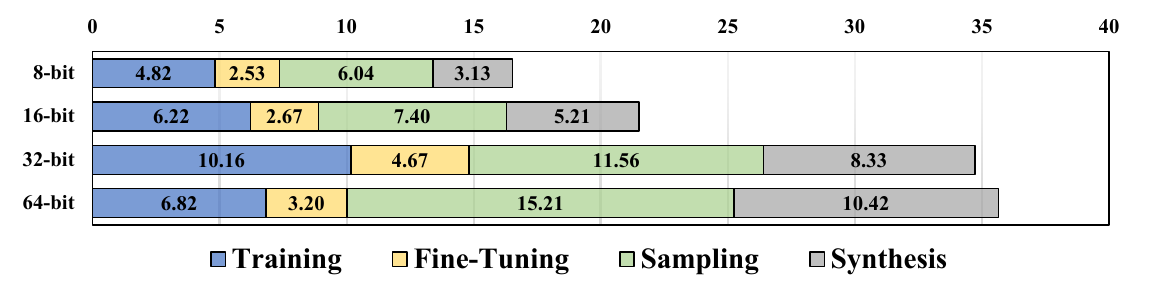}
  \caption{Runtime of multiplier optimization (time unit: h).}
  \label{fig:runtime}
\end{figure}

\noindent \textbf{Self-Reflection.}
To demonstrate that self-reflection enables effective gradient-based guidance, TABLE~\ref{tab:legalization} shows the legalization results for compressor trees generated with different methods.
The unconditional diffusion model attempts to interpolate across distinct circuit structures, which may lead to substantial DRVs.
For conditional models with conservative guidance strength, the discrepancy between predicted and target QoR cannot be effectively minimized, resulting in similar DRVs as the unconditional counterparts.
By incorporating self-reflection to minimize $\mathcal{L}(y^*, \tilde{y})$, \papername{} restricts the diffusion model to interpolate on fewer targeted circuit structures. Consequently, the generated designs exhibit fewer DRVs and can be easily legalized.

\noindent \textbf{Model Fine-Tuning.}
Fig.~\ref{fig:impact_finetune} shows the QoR optimization progress with iterative model fine-tuning.
By enhancing the models with sampled high-quality designs, \papername{} achieves focused exploration near the Pareto frontier, with increased probability of generating high-potential circuit designs. As a result, \papername{} progressively improves multiplier QoR, until converging at an optimized state.

\subsection{Discussions}
\label{sec:discussions}

\noindent \textbf{Runtime Analysis.}
Fig.~\ref{fig:runtime} shows the runtime breakdown for comprehensive optimization of multipliers, including CT and CPA. The total runtime consists of four parts: (1) training diffusion models and cost predictors; (2) fine-tuning the models; (3) conditional sampling for new CT and CPA design; (4) QoR evaluation by running EDA tools. 
Unlike many RL-based approaches~\cite{roy2021prefixrl,zuo2023rl,lai2024scalable}, \papername{} enables parallel synthesis on sampled designs, which significantly alleviates the bottleneck of EDA tool invocations.
Although model training and design generation take comparable amount of time to synthesis in our current setup, both tasks exhibit high parallelism and can be accelerated proportionally with the number of GPUs employed.

\noindent \textbf{Selection of Generative Models.}
In \papername{}, we adopt diffusion model over alternative generative approaches due to its high generation quality and high controllability. We observe that these advantages are crucial for achieving efficient optimization. As shown in TABLE~\ref{tab:legalization}, we also apply VAEs~\cite{kingma2013auto} to generate CT designs with complex design constraints (Equation~\eqref{eq:ct_drc_nonzero}-\eqref{eq:ct_drc_final}).
While VAEs have been adopted for prefix adder optimization in previous work~\cite{song2024circuitvae}, they perform poorly in generating correct CT designs. The produced CTs take excessive steps to legalize and even become irreparable.
This suggests that the deficiency of VAEs in capturing fine-grained details~\cite{gulrajani2016pixelvae} renders them unsuitable for optimizing complex arithmetic circuits.

\section{conclusion}
\label{sec:conclusion}

This work introduces \papername{}, an arithmetic circuit optimization framework enabled with conditional diffusion models.
By introducing conditional design generation process and model fine-tuning techniques, \papername{} consistently explores promising design candidates, enabling efficient design optimization. 
Future work will address the optimization for variable-width arithmetic components and explore extending the methodology to other components and broader applications.

\section*{Acknowledgement}
This work is supported in part by Beijing Natural Science Foundation (Grant No. L243001), National Natural Science Foundation of China (Grant No. 62032001, 62034007), National Key Research and Development Program of China (Grant No.2023YFB4402204,2021ZD0114702), 111 Project (B18001), the Hong Kong Research Grants Council (RGC) under Grant No. 14212422, 14202824, and C6003-24Y.

{
\bibliographystyle{IEEEtran}
\bibliography{./main}
}

\end{document}